\documentstyle[12pt]{article}

\begin{document}
\newcommand {\be}{\begin{equation}}
\newcommand {\ee}{\end{equation}}
\newcommand {\bea}{\begin{array}}
\newcommand {\cl}{\centerline}
\newcommand {\eea}{\end{array}}
\renewcommand {\thefootnote}{\fnsymbol{footnote}}
\baselineskip 0.65 cm
\begin{flushright}
IPM/P-99/04\\
hep-th/9901080
\end{flushright}
\begin{center}
{\Large{\bf Open Strings in a B-field Background as Electric Dipoles }}
\vskip .5cm

 M.M. Sheikh-Jabbari
\footnote{ E-mail:jabbari@theory.ipm.ac.ir } \\

\vskip .5cm

 {\it Institute for studies in theoretical Physics and Mathematics 
IPM,

 P.O.Box 19395-5531, Tehran, Iran}\\
\end{center}

\vskip 2cm
\begin{abstract}
Studying dynamics of open strings attached to a D2-brane in a NS two
form field background, we find that these open strings act as {\it dipoles}
of $U(1)$ gauge field of the brane. This provides an string theoretic
description of the flux modifications needed for the DBI action on
noncommutative torus.

\end{abstract}
\newpage
{\it Introduction}
\newline
After the interesting work of A. Connes, M. Douglas and A. Schwarz (CDS) [1],  
noncommutative geometry and especially the noncommutative torus 
is shown to play a crucial role in the M-theory compactifications [2, 3, 4, 5,  
6, 7, 8, 9, 10, 11, 12, 13, 14, 15].
In [1], CDS conjectured that the Super Yang-Mills (SYM) formulated on a 
noncommutative torus describes the discrete light cone quantization of M-theory 
when we have a non-zero three form background of 11 dimensional supergravity. In that paper,
CDS used the SYM action with an additional topological term proportional to
the field strength. This topological term in the case of noncommutative
compact spaces leads to some modification in BPS spectrum, necessitated by 
the U-duality.

Since the eleven dimensional three-form is related to the NSNS two form field
of string theory, the CDS conjecture means that the low energy dynamics of 
D0-branes or more generally any D-brane, in a B field background is described 
by a gauge theory on noncommutative torus, where $\theta$, the deformation   
parameter of the torus is determined by the B field background [2, 3, 4, 5, 6, 
7, 8, 12, 14, 15]. It was argued in [8] that, there should be an additional 
term in the Matrix model action proportional to $\dot{X}$, the time derivative 
of the M(atrix)-valued collective coordinates of D0-branes, $X$. It is easy 
to show that this term can be obtained from the topological term discussed 
earlier by CDS, but as they mentioned, they had no physical reasoning for 
adding this term. Also there were argued that the coefficient of this term is 
proportional to the winding number of longitudinal membranes.

Generalizing the ideas of [2], it was shown in [14] that the scattering 
amplitudes for open strings attached to a D2-brane in the B field background 
in the low energy limit,is properly described by the SYM defined on a  
noncommutative two torus with deformation parameter identified with the 
background B field. It was also discussed in [14] that higher
$\alpha'$ corrections to the open string scattering amplitude, are given
by the DBI action defined on noncommutative torus. 

In this paper, studying the open string dynamics, in a non-zero electric field 
of the D2-brane, we show that in the case 
of non-zero B-field background these open strings act as an {\it electric
dipole}. Hence in a non-vanishing electric field one should {\it modify}
the SYM action, the action describing the open strings dynamics at low energies, 
by adding a term proportional to dipole moment of these open strings, which 
shows the interaction of these dipoles with electric field. 
We will argue that, in the noncommutative gauge theory, as argued by
Hofman and Verlinde [13],
generalizing the usual  $Tr$ of non-Abelian gauge theory to $Tr_{\theta}$, 
defined by CDS, the Sl(3,Z)$\times$Sl(2,Z) symmetry still remains and it 
{\it automatically} reproduces the term, we find it as dipole interactions 
in string theory. However if we do the quantization procedure correctly [16], 
unlike the works of [1,8], there is no need to add an additional term 
proportional to $\dot{X}$ to the M(atrix)-model action compactified on the 
noncommutative torus.
In this way we provide a more intuitive description of using 
noncommutative geometry methods. 

Elaborating on this point we will show that, considering the dipole-background 
electric interactions, assures the full U-duality group, from the string 
theoretic calculations, {\it without} using the noncommutative geometry.
From the M-theory point of view,
the dipole moment which are {\it conserved} quantities give the winding of  
longitudinal membrane [1,8]. This winding number is a new degree of freedom 
should be added to the usual Matrix model, when we have a non-vanishing three
form field of the 11 dimensional supergravity.

{\it Open String Dipoles}
\newline
It was shown and discussed by Witten [17], that the massless states of open 
strings attached to a $D_p$-brane, form a vector multiplet of a $N=1$ $D=10$ 
$U(1)$ gauge theory dimensionally reduced to (p+1)\footnote{Unless it is 
mentioned explicitly by gauge theory here we mean SYM or DBI where either of 
which are gauge invariant.}. There, it was also discussed that the end point
of any open string attached to a D-brane carries the unit charge of that $U(1)$
gauge theory. This point was explicitly worked out  by studying the BPS 
excitation of the $U(1)$ SYM or DBI [18].
Let us consider an open string having its both ends on the same brane. Since 
these open strings are oriented, ends of them, from the gauge theory point of  
view, look like plus-minus unit charges. So at the first sight, it seems 
that they form an electric dipole and not a gauge particle of that $U(1)$
theory. This problem is easily resolved if we look at the open strings more 
carefully. The bosonic part of the mode expansion of such vector states
are:
\be
X^{\mu}=x^{\mu}+ p^{\mu}\tau+ \sum_{n\neq 0}a^{\mu}_n{e^{-in\tau} \over n}
\cos n\sigma \;\;\;\; \mu=0,...,p.
\ee
Since the vector state is described by $b_{-1/2}^{\mu}|VAC>$\footnote{ 
$b_{-1/2}^{\mu}$ is the NS sector creation operator.}, $X^{\mu}(0,\tau)$
and $X^{\mu}(\pi,\tau)$ for such states have the same value. Hence the two
end points, or the plus-minus charges, are really on top of each other and 
there is no dipole moment, and these open strings simply give the gauge 
multiplet of $U(1)$ gauge theory.
In order to discuss the same issue in the case with non-zero B-field,
first one should build the mode expansion of corresponding open strings.
Here after for simplicity, we only consider a D2-brane with non-zero B-field
on it, however our discussion can easily be generalized to the case of other
D-branes. The open strings attached to such a D2-brane are described by
[18] 

\be
\left\{  \begin{array}{cc}
\partial_{\sigma}X^0=0 \\
\partial_{\sigma}X^{i}+B^i_j \partial_{\tau}X^j=0  \;\;\; i,j=1,2\\
\partial_{\tau} X^a=0 \;\;\;\ ,\;\;\; a=3,...,9.
\end{array}\right.
\ee
Hence the mode expansion of these open strings are [14]
\be
\left\{  \begin{array}{cc}
X^0=x^0+ p^0\tau+ \sum_{n\neq 0}a^0_n{e^{-in\tau} \over n}\cos n\sigma\\
X^{i}=x^i+(p^i \tau-B^i_j p^j \sigma)+ \sum_{n\neq 0} {e^{-in\tau} \over n}
\bigl(ia^i_n \cos n\sigma + B^i_j a^j_n \sin n\sigma \bigr) \;\;\;\ i,j=1,2\\
X^a=x^a+ \sum_{n\neq 0}a^a_n {e^{-in\tau} \over n}\sin n\sigma ,
\end{array}\right.
\ee
where $x^i$ show an arbitrary point on D2-brane. 

Let us again study these open strings from the gauge theory point of view. In 
this case, unlike the previous case ($B=0$) the plus-minus charges 
locating at the end points of open strings are not coincident any more,
hence these open strings look like {\it electric dipoles} with the moment 
${\cal P}^i$:
\be
{\cal P}^i=X^{i}(0, \tau)-X^{i}(\pi, \tau)=\pi B^i_j p^j.
\ee
As we see the dipole moment is proportional to $B$ field and open string
momenta, and is always perpendicular to the momentum vector, $p^i$. 

As it is discussed by many people, the gauge theory governing the D-brane
or open strings dynamics in a non-zero B field background, is not a usual
DBI action but, it is the DBI  
defined on a noncommutative torus. In the noncommutative case again we
can talk about the electric charges of the $U(1)$ theory defined by
the {\it zero momentum sector} of open strings. In the $B=0$ case (or
more generally any rational B), since
our theory enjoys the (2+1) Lorentz invariance, we can always make these
dipoles to be zero. In contrast, for $B \neq 0$ our theory suffers from the
lack of Lorentz symmetry [1, 12] then, especially when the brane is
compact, these dipoles can not be removed and they are the intuitive
origin of the
noncommutativity of torus and the Moyal bracket structure.
Form the gauge theory point of view,  these dipoles are really the gauge 
particles of the noncommutative gauge theory, where they interact through the 
Moyal bracket terms of the action. And one can also understand the noncommutative 
SYM, either by the open string dynamics [14], or by these dipole-dipole 
interactions.

Here we briefly discuss some of the issues of the dipole description of 
noncommutative gauge theory in (2+1) dimensions. A more extensive work
will appear [19].

{\it i}) Dipole moment conservation; Since the dipole moment is proportional
to the open string momentum, the momentum conservation in each vertex will
immediately result in the dipole moment conservation.

{\it ii}) Dipoles always move so that their dipole moment are normal to their
momentum.

{\it iii}) Because of Moyal bracket structure, parallel dipoles are 
non-interacting.

{\it iv}) The high energy dipole-dipole scattering is suppressed by the  
Moyal bracket structure.

{\it Interaction of Dipoles with Electric Background}
\newline
Besides the dipole-dipole interactions, which are described by the Moyal 
gauge theory, i.e. the gauge theory defined on a noncommutative torus, there
are dipole-electric background interactions, which are not present in the 
noncommutative gauge theory and we should add them.
In other words, to have a theory fully invariant under U-duality, 
the $Tr\rightarrow Tr_{\theta}$ substitution should be done not only   
for free fields, but also for currents and fluxes (or BPS charges) [13]. 

To work out the form of the dipole-background interactions, first one should
discuss the explicit form of dipole moment. So let us consider a (D2-D0)-brane
system, winding around the two torus defined by 
\be
\tau={R_2 \over R_1}e^{i\alpha} \;\;\;\; , \;\;\;\;\ \rho=iR_1R_2\sin\alpha+B.
\ee
Along the calculations of [12], mode expansion of open strings attached to 
such a brane system is given by eq. (3), where $p^i$ in the usual complex
notation of the torus, is
\be
p={(r_1+q_1\tau)(n+m\rho) \over |n+m\rho|^2}\sqrt{{\rho_2 \over \tau_2}}
\ee

with $r_1,q_1$ being two arbitrary integers, (n,m) two integers that their 
greatest common divisor shows the winding number of D2-brane around the
torus and their ratio, ${m\over n}$ gives the density of D0-branes distributed
on the D2-brane. Hence according to eq. (4):
\be
{\cal P}=\pi iB{(r_1+q_1\tau)(n+m\rho)\over |n+m\rho|^2}
\sqrt{{\rho_2\over \tau_2}}
\ee
To include the dipole-background interactions, we should add the proper  term
to the gauge theory action. Since we finally want to find the BPS spectrum of 
the the (D2-D0)-brane system, we use the DBI action:
\be
S=S_0+\int d^3x {\cal P}^i {\cal F}_{0i}=\int L_{mod.}dt,
\ee
where $S_0$ is the usual DBI action:
\be
S_0={-1\over g_s}\int d^3x \sqrt{det(g+{\cal F})}=\int L_0 dt.
\ee
Here we have put $l_s=1$ but at last will reintroduce it. 
There could also be a WZ term, $\int C\wedge {\cal F}$, since it does not
alter our arguments much, we will not consider it at this stage, but will 
come back to it in our final results.
To check the BPS spectrum, we build the Hamiltonian:
\be\bea{cc}
H={\cal F}_{0i}{\partial L_{mod.} \over \partial{\cal F}_{0i}}-L_{mod.} \\
\;\;={1\over l_sg_s}{|n+\rho m|\over \sqrt{\rho_2}}(1+|\Pi_0|^2g_s^2)^{1/2},
\eea.
\ee
with
\be
\Pi_0={\partial L_{0} \over \partial{\cal F}_{0i}}=
{\partial L_{mod.} \over \partial{\cal F}_{0i}}-{\cal P}.
\ee
The conjugate momenta of ${\cal F}_{0i}$, 
${\partial L_{mod.} \over\partial{\cal F}_{0i}}$, should be quantized as dual
torus vector. Hence comparing eq. (11) with the results of [12], every thing is the
same except for the shift by ${\cal P}$ in the conjugate momenta:
\be
r_2,\; q_2 \rightarrow r_2+r_1\rho,\; q_2+q_1\rho.
\ee
Finally putting all of these together, we find  
\be
H={1\over l_sg_s}{|n+m\rho|\over \sqrt{\rho_2}}\bigg(1+g_s^2{1 \over\tau_2} 
{|(r_2+r_1\rho)+\tau(q_2+q_1\rho)|^2\over|n+m\rho|^2}\bigg)^{1/2}.
\ee
If we had also considered the WZ term along the lines of [13], would end up
with:
$$
H={1\over l_sg_s}{|n+m\rho|\over \sqrt{\rho_2}}\bigg(1+g_s^2{1 \over\tau_2} 
\big(|{(r_2+r_1\rho)+\tau(q_2+q_1\rho)\over|n+m\rho|^2}+(C_2+C_1\tau)|^2\big)
\bigg)^{1/2}.
$$
This spectrum as discussed in [13] has both Sl(3,Z) and $Sl(2,Z)_N$ symmetries.
In order to compare our results with those of Matrix model [1,8], one should
take the  $l_s, g_s \rightarrow 0$ limit, which again will give the results of [12],
with:
\be
r_2,\; q_2 \rightarrow  r_2-r_1\theta,\; q_2-q_1\theta.
\ee

{\it M-theory Interpretations} 
\newline
As we see, considering the missing dipole-background interactions, 
modifies conjugate momentum of the electric fields living on the two torus
by adding a term proportional 
to open strings momentum, eq. (4). In the M-theory side, 
the open strings momenta and m, number of the D0-branes, 
are conserved  charges related to $C_{\mu ij}$ components of 11 dimensional 
three-form, where $\mu=0,..,7$ and $i,j$ denoting the directions of $T^3$
M-theory compactified on. 
And $r_2,q_2$, n (roughly speaking, the number of D2-branes), are related
to  
$g_{\mu i}$ components of metric. More precisely $(r_1,q_1,m)$ and 
$(r_2,q_2,n)$ form two $Sl(3,Z)$ vectors, and $(r_2,r_1)$,$(q_2,q_1)$, 
$(n,m)$ act as $Sl(2,Z)_N$ doublets. In terms of M2-branes these conserved 
charges are the KK momenta and winding modes of M2-brane compactified on
$T^3$. Generalizing the ideas of [21], to the non-zero $C_{ijk}$, we  
find that the momenta should be modified by windings [1], similar to what
we have found here from the string theory arguments.

{\it Concluding Remarks}
\newline
In this work completing the results of [12,14], we provide a string theoretic
justification of the $Tr\rightarrow Tr_{\theta}$ substitution, when we 
have a NSNS two-form background field. It was argued by Hofman and Verlinde 
[13] that a proper Matrix model treatment of M-theory with the three-form
background is given by Born-Infeld on noncommutative torus. It was also 
discussed [13] that, going from $T^2$ to $T^2_{\theta}$, one should modify 
the related {\it fluxes} too.
Here we studied the open strings attached to D2-brane with a non-zero B field
more carefully and in this way provide an string theoretic description of 
the flux modification, and observed that these open strings look like electric
dipoles of $U(1)$ gauge theory. Considering these dipoles sheds
light on the noncommutative geometry methods and shows an explicit way to 
study the renomalizability of noncommutative gauge theories [20].

{\bf Acknowledgements}

I would like to thank H. Arfaei for many  fruitful discussions.
I also thank Pei-Ming Ho for comments.

\end{document}